\documentclass[aps,twocolumn,PRB,reprint,superscriptaddress]{revtex4-1}
\usepackage{amsmath}
\usepackage{epsfig}
\usepackage{graphicx}
\usepackage{graphicx, color, epstopdf}
\usepackage{bm}
\usepackage{amssymb}
\usepackage{hyperref}
\usepackage{xcolor}
\usepackage{subfigure}
\begin{document}

\title{Random-gate-voltage induced Al'tshuler-Aronov-Spivak effect in topological edge states}
\author{Kun Luo}
\affiliation{National Laboratory of Solid State Microstructures and department of Physics, Nanjing University, Nanjing, 210093, China}
\affiliation{Collaborative Innovation Center of Advanced Microstructures, Nanjing University, Nanjing 210093, China}

\author{Wei Chen}
\email{Corresponding author: pchenweis@gmail.com}
\affiliation{National Laboratory of Solid State Microstructures and department of Physics, Nanjing University, Nanjing, 210093, China}
\affiliation{Collaborative Innovation Center of Advanced Microstructures, Nanjing University, Nanjing 210093, China}

\author{Li Sheng}
\affiliation{National Laboratory of Solid State Microstructures and department of Physics, Nanjing University, Nanjing, 210093, China}
\affiliation{Collaborative Innovation Center of Advanced Microstructures, Nanjing University, Nanjing 210093, China}

\author{D. Y. Xing}
\affiliation{National Laboratory of Solid State Microstructures and department of Physics, Nanjing University, Nanjing, 210093, China}
\affiliation{Collaborative Innovation Center of Advanced Microstructures, Nanjing University, Nanjing 210093, China}
\begin{abstract}
Helical edge states are the hallmark of the quantum spin Hall insulator.
Recently, several experiments have observed
transport signatures contributed by trivial edge states,
making it difficult to distinguish between the topologically trivial and nontrivial phases.
Here, we show that helical edge states can be
identified by the random-gate-voltage induced
$\Phi_0/2$-period oscillation of the averaged electron
return probability
in the interferometer constructed by the edge states.
The random gate voltage can highlight the $\Phi_0/2$-period
Al'tshuler-Aronov-Spivak oscillation proportional to $\sin^2(2\pi\Phi/\Phi_0)$ by quenching
the $\Phi_0$-period Aharonov-Bohm oscillation.
It is found that the helical spin texture induced $\pi$ Berry phase is key to such
weak antilocalization behavior with zero return probability at $\Phi=0$.
In contrast, the oscillation for the trivial edge states may exhibit either weak localization or
antilocalization depending on the strength of the spin-orbit coupling, which have
finite return probability at $\Phi=0$.
Our results provide an effective way for the identification of the helical edge states. The
predicted signature is stabilized by the time-reversal symmetry
so that it is robust against disorder and does not require any fine adjustment of system.
\end{abstract}

\date{\today}

\maketitle

\section{INTRODUCTION}
Over the past decade, topological phases of matter
such as topological insulator and superconductor have
become a new research field of condensed matter physics \cite{RevModPhys.83.1057,RevModPhys.82.3045}.
As the first theoretically proposed and experimentally implemented topological
material, the quantum spin Hall insulator (QSHI)
has attracted broad research interest \cite{Kane05prl,Bernevig06prl,Bernevig1757,Liu08prl,Konig766,Knez11prl,Knez14prl,Du15prl}.
The fingerprint of QSHI is the existence of topologically
protected helical edge states
at the sample boundary \cite{Kane05prl,Bernevig1757}.
Due to spin-momentum locking, helical edge states
provide an interesting platform to study exotic electronic properties.
Moreover, such strongly spin-orbit coupled
system can also lead to important applications such as
spintronic device, low-consumption
transistor \cite{RevModPhys.83.1057,RevModPhys.82.3045} and topological
quantum computation \cite{Fu08prl,Fu09prb,Kitaev01pu}.

\begin{figure}
\centering
\includegraphics[width=0.9\columnwidth]{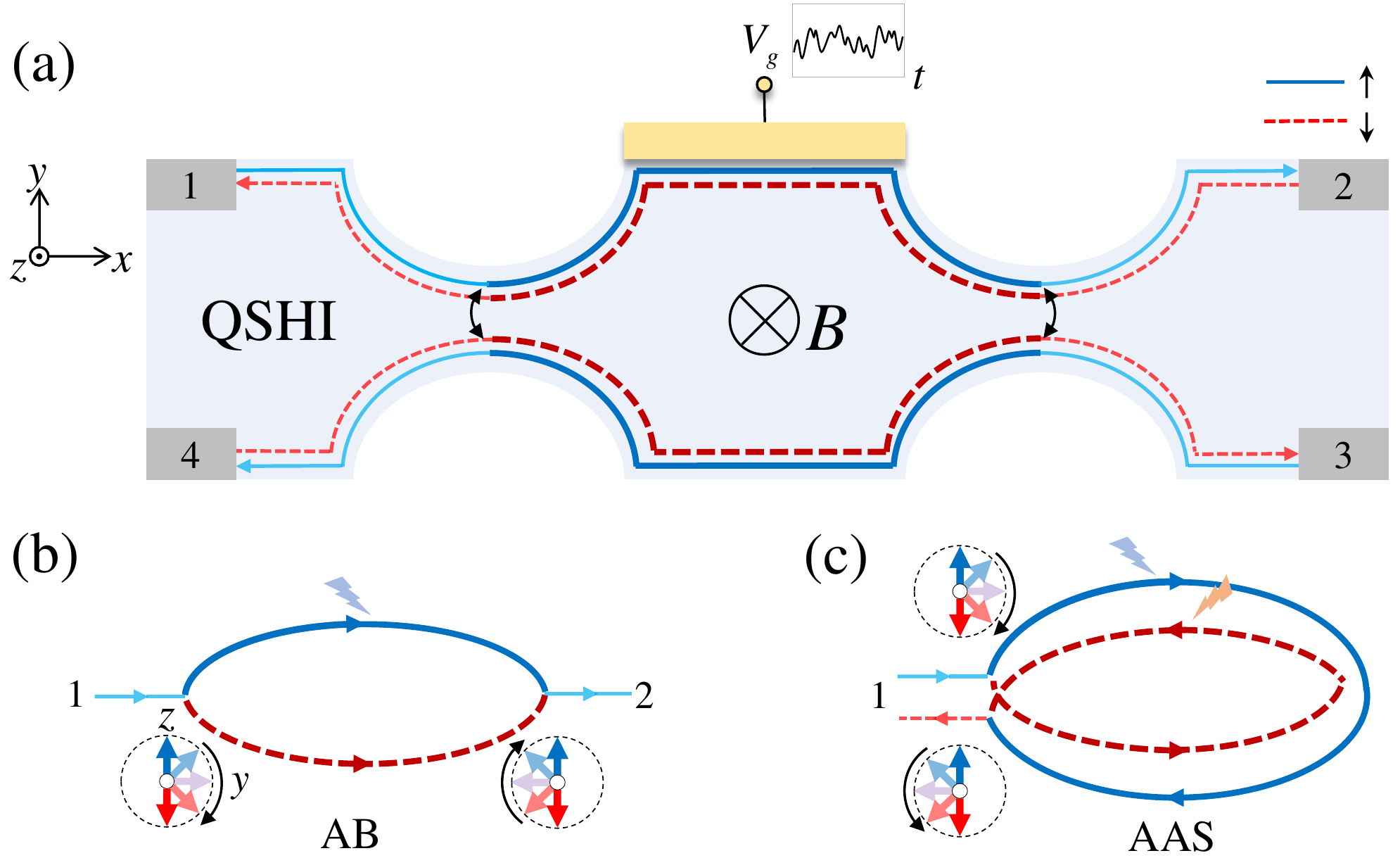}\\
\caption{(a) The interferometer
constructed by the edge states of the QSHI.
The spin up (down) edge states are denoted
as blue solid (red dashed) lines,
with their moving directions labeled by arrows.
Two point contact structures allow the coupling between
the upper and lower edge states. A side gate (yellow bar) with random voltage $V_g$ is
deposited at the upper edge to introduce dephasing effect.
Weak magnetic field $B$ is imposed perpendicular to the sample.
Four terminals (1-4) are sketched as gray bars.
(b) The AB interfering loop is fragile to the random gate.
The spin flipping accompanies the inter-edge transmission.
(c) AAS interfering loop is immune to the random gate. The
$\pi$ Berry phase associated with the spin rotation
gives rise to zero return probability.
}\label{fig1}
\end{figure}

Although the edge states have been confirmed by
measuring the edge conductance in various
QSHI materials \cite{Konig766,Knez11prl,Knez14prl,Du15prl},
its helical nature remains to be identified.
Recently, topologically trivial edge states induced by band bending
have been observed in various materials, such
as InAs/GaSb quantum wells \cite{Nguyen16prl,Nichele16njp,Mueller17prb,Sazgari19prb},
InAs \cite{Mueller17prb,Vries18prl} and InSb \cite{Vries19prr}.
These trivial edge states result in similar
transport signatures which make it difficult to
discriminate them from the helical ones.
Several theoretical proposals have been put forward for the detection
of helical spin texture of the edge states \cite{Hou09prl,Schmidt11prl,Das11prl,Soori12prb,Edge13prl,Romeo14prb,Chen16prl},
yet their implementation remains challenging,
which are either beyond state-of-the-art fabrication technique or
requires a fine tuning of the sample parameters.
Therefore, it is highly demanded to find a clear and robust signature
to discriminate the helical edge states from the
trivial ones without fine tuning of the system.

In this work, we investigate coherent electron
transport through an Aharonov-Bohm (AB) interferometer
composed of edge states; see Fig. \ref{fig1}(a).
By imposing a random gate to the interferometer,
the main $\Phi_0$-period AB oscillation is suppressed
due to the dephasing effect [cf. Fig. \ref{fig1}(b)], giving rise to
a dominant Al'tshuldr-Aronov-Spivak (AAS) oscillation
with $\Phi_0/2$ periodicity [cf. Fig. \ref{fig1}(c)].
Different from the conventional AAS oscillation,
here, the weak antilocalization
effect is induced by the random gate rather than
disorder averaging \cite{al1981aaronov,al1982observation}.
The results are generally stabilized by the time-reversal symmetry,
thus offering a robust signature of the
helical spin texture that does not rely much on the
sample details.
Specifically, the AAS oscillation of the return probability
takes the form $\bar{R}=\mathcal{R}\sin^2(2\pi\Phi/\Phi_0)$.
The absence of return probability at $\Phi=0$ is attributed to the
destructive AAS interference induced by the $\pi$ spin Berry phase
accumulated during inter-edge tunneling, a direct
result of the helical spin texture.
For the trivial edge state,
the return probability take the general form $\bar{R}=\pm\mathcal{R}\sin^2(2\pi\Phi/\Phi_0)+\mathcal{R}_0$
with $\mathcal{R}_0>0$, where the ``$\pm$'' is determined by the strength of the spin-orbit coupling.
In contrast to the helical edge states,
nonzero return probability occurs at $\Phi=0$
for the trivial edge states
due to the doubling of transport channels.
The different oscillation patterns of
the return probability can be measured
by the differential conductance.
Therefore,
our scheme provides an unambiguous evidence
to discriminate between the nontrivial and trivial edge states.

The rest of this paper is organized as follows. In Sec. \ref{pp},
we explicate the basic physical picture of the
main results. In Sec. \ref{sm}, we adopt a scattering matrix
analysis based on the effective model of the edge states
to solve the random gate induced $\Phi_0/2$-period
oscillation of the return probability. Detailed numerical
simulations on the lattice model is conducted in Sec. \ref{lm} to
show the robustness of the main results for the helical edge states.
In Sec. \ref{te}, we study the AAS oscillation for the trivial edge states
and compare the results with that of the helical edge states.
Finally, a brief summary and discussion are given in Sec. \ref{sd}.

\section{Physical picture}\label{pp}
The AB interferometer constructed by the edge states of the QSHI
is illustrated in Fig. \ref{fig1}(a), where the spin up (down) edge states
propagate in the clockwise (anticlockwise) direction. Two point contacts (PCs)
are fabricated to introduce inter-edge coupling \cite{strunz2020interacting}, such that
the central region of the setup forms a closed interfering loop.
A weak magnetic field is applied, which acts on the electron motion through
the flux modulation $\Phi=BS$ in the area $S$ enclosed by the interference loop.
At two PCs,
Rashba spin-orbit coupling can be induced by local electric
fields \cite{rothe2010fingerprint,strom2010edge,vayrynen2011electrical,strunz2020interacting},
which gives rise to inter-edge transmission, i.e., electrons in one edge
transmit to the opposite edge with the same moving direction
accompanied by a spin flipping \cite{vayrynen2011electrical,Chen16prl}.
Here, we apply a gate voltage $V_g$ to the upper edge; see Fig. \ref{fig1}(a),
which introduces an additional phase factor $\delta=eV_gL/v$
to the wave function of the upper electron, with $L$ the length of the gate
and $v$ the velocity of the edge states.
Here, $V_g$ takes random values,
which is key to the main results of this work.

Consider an electron propagates coherently in the AB interferometer,
which contains multiple interfering loops.
Two typical ones among them are the main AB and AAS interfering loops,
as shown in Figs. \ref{fig1}(b) and \ref{fig1}(c), respectively.
The main AB loop is composed of
the upper and lower transmission channels, which results in
a $\Phi_0$-period oscillation of transmission probability.
Such interference is sensitive to the phase shift $\delta$
induced by $V_g$
so that a random $V_g$ causes dephasing effect
and quenches such AB oscillation.
In contrast, the AAS interference
consists of two backscattering
trajectories being the time-reversal
counterpart to each other.
As the time scale of $V_g$ fluctuation is
much larger than that an electron spends
inside the interferometer, the pair of paths gain the same random phase $\delta$.
Therefore, such AAS interference survives,
giving rise to a dominant $\Phi_0/2$-period oscillation
of the return probability.

Importantly, the electron return probability
contains the key information of the helical
spin texture of the edge states.
Consider an electron incident from terminal 1 in Fig. \ref{fig1}(a),
both two backscattering
trajectories involves an inter-edge transmission
along with a $\pi$ spin rotation about the $x$ axis
in opposite directions denoted by the rotation operator
$O_x({\pm\pi})$; see Fig. \ref{fig1}(c).
Accordingly, the backscattering wave function
takes the form $r_0e^{i\delta}[O_x(-\pi)+e^{i4\pi\Phi/\Phi_0}O_x(\pi)]|\uparrow\rangle$,
with $r_0$ the backscattering amplitude for each path.
It reduces to $r_0e^{i\delta}[1+e^{i4\pi\Phi/\Phi_0}O_x(2\pi)]|\downarrow\rangle$
and gives rise to the probability $\propto\sin^2(2\pi\Phi/\Phi_0)$
using the property of the spin Berry phase $O_x(2\pi)=e^{i\pi}$.
Since the spin rotation is enforced by the spin-momentum locking,
such an oscillation behavior directly reveals the spin texture of the edge states.
For zero magnetic field with $\Phi=0$, the return probability vanishes,
which is prohibited by the time-reversal symmetry of the helical
edge states. A small magnetic flux breaks
the time-reversal symmetry and increases the
return probability, indicating a weak antilocalization effect.

\section{Scattering matrix analysis}\label{sm}
In this section, we study electron transport in the interferometer
using the scattering matrix approach based on
the low-energy model of the edge states.
The scattering matrix of the whole interferometer
is obtained by combining the scattering
matrices at two
PCs and involving the phase accumulation
during electron propagation.
The scattering matrix at the left PC can be parameterized as
\begin{equation}\label{S}
S_L=\left(
    \begin{array}{cccc}
      0 & r_1 & -t_1 & -t_2 \\
      -r_1 & 0 & -t_2 & -t_1 \\
      t_1 & t_2 & 0 & r_2 \\
      t_2 & t_1 & -r_2 & 0 \\
    \end{array}
  \right),
\end{equation}
which relates the incident and outgoing waves via $b_L=S_La_L$.
The four components of the wave $a_L, b_L$ correspond to
the upper and lower channels on the left and right side of the PC.
Here, the antisymmetric condition $S_L^T=-S_L$ is taken into account
due to the time-reversal symmetry of the helical states
\cite{bardarson2013quantum}.
The amplitudes $t_1, t_2$ are the intra- and inter-edge
transmission, respectively; $r_1, r_2$ are the inter-edge
reflection on both sides of the PC,
while the intra-edge reflection with spin flipping is prohibited by the time-reversal symmetry.
The matrix for the right PC $S_R$
is defined in the same way. The phase modulation of the interferometer
comes from the gate voltage $V_g$ deposited on the upper
channel [cf. Fig. \ref{fig1}(a)] and the flux $\Phi$
in the middle area encircled by the interfering loop,
which can be described by the matrix
\begin{equation}
S_M=\left(
  \begin{array}{cccc}
    0 & 0 & e^{i(\delta-\phi_1)} & 0 \\
    0 & 0 & 0 & e^{i\phi_2} \\
    e^{i(\delta+\phi_1)} & 0 & 0 & 0 \\
    0 & e^{-i\phi_2} & 0 & 0 \\
  \end{array}
\right)
\end{equation}
where $\phi_{1,2}$ are the extra phase due to the magnetic field,
with the sum $\phi=\phi_1+\phi_2=2\pi\Phi/\Phi_0$
being gauge invariant.

Combining three matrices $S_L, S_M, S_R$
yields the scattering matrix for the whole interferometer.
In the following, $S_R=S_L$ is adopted for simplicity,
which will not change the main results.
For an electron injected from terminal 1,
its return probability $R$ and transmission probability $T$ to terminal 2
are obtained after some algebra as
\begin{equation}\label{RT}
\begin{split}
R&=F^{-1}4|r_1t_1t_2|^2\sin^2\phi\\
T&=F^{-1}\Big[|X|^2+|Y|^2+|Z|^2+2|XY|\cos(\phi-\delta+\theta_{xy})\\
&\ \ \ +2|XZ|\cos(\phi+\delta+\theta_{xz})+2|YZ|\cos(2\delta+\theta_{yz})\Big]\\
F&=1+K^4+2K^2\big[\cos2(\delta+\nu)+2\cos^2\phi\big]\\
&\ \ \ +4K(1+K^2)\cos\phi\cos(\delta+\nu),
\end{split}
\end{equation}
where $X=t_1^2+t_2^2r_1r_2$, $Y=t_1^2r_1r_2$, $Z=t_2^2$, $\theta_{ij}=\theta_i-\theta_j$
with $\theta_{i=x,y,z}$ are the arguments of $X, Y, Z$, respectively, $K=|r_1r_2|$ and $\nu=\text{arg}(r_1r_2)$.

The scattering probabilities in Eq. \eqref{RT}
contains the information of
the competition between the AB and AAS effects.
For a fixed $V_g$ or equivalently, the phase
factor $\delta$, the transmission $T$
is dominated by the AB oscillation with $\Phi_0$ periodicity.
Due to the current conservation, the $\Phi_0$-period component
also appears in the oscillating pattern of the return probability $R$.
As $\delta$ takes
random values or equivalently, it is integrated out,
the first-order AB oscillation quenches and the
$\Phi_0/2$-period AAS oscillation dominates
the transport.
This fact can be clearly seen
in the weak reflection limit
of the PCs, that is $K\ll1$.
By expanding $F$ in Eq. \eqref{RT} to the first order of $K$,
the return probability reduces to
\begin{equation}
R\simeq 4|r_1t_1t_2|^2\sin^2\phi\big[1-4K\cos\phi\cos(\delta+\nu)\big],
\end{equation}
which contains both $\Phi_0$ and $\Phi_0/2$ periodicity.
To see the effect due to the random gate voltage,
we integrate out $\delta$, which yields
\begin{equation}
\bar{R}\simeq4|r_1t_1t_2|^2\sin^2\phi,
\end{equation}
where only the $\Phi_0/2$-period oscillation remains.
Similarly, the averaged transmission probability
reduces to
\begin{equation}
\begin{split}
\bar{T}&=|X|^2+|Y|^2+|Z|^2\\
&+2K|XY|\big[\cos(2\phi+\theta_{xy}+\nu)+\cos(\theta_{xy}+\nu)\big]\\
&+2K|XZ|\big[\cos(2\phi+\theta_{xz}-\nu)+\cos(\theta_{xz}-\nu)\big],
\end{split}
\end{equation}
in which only the $\Phi_0/2$-period oscillation remains, the same as $\bar{R}$.
The random $V_g$ introduces dephasing
effect to the AB interference process while
retains phase coherence between
paired AAS interference loops.
As a result, the AAS effect dominates the transport,
giving rise to the $\Phi_0/2$-period oscillation.

\begin{figure}
\centering
\includegraphics[width=\columnwidth]{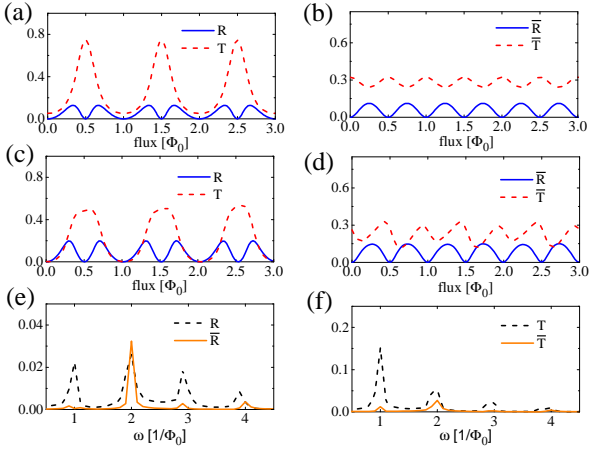}\\
\caption{The oscillation pattern of $R, T$ ($\bar{R}, \bar{T}$) from the edge state
analysis for (a) $\delta=0$ and (b) random $\delta$.
The matrix elements in Eq. \eqref{S}
are set as $r_1=\sqrt{0.3}$,
$t_1=\sqrt{0.5}i$ and $t_2=\sqrt{0.2}$.
The results of the lattice simulation for (c) $V_g=0$ and (d)
random $V_g$ with $\mathcal{W}=10$ meV and an incident energy of 1.5 meV.
The FFT spectrum of (e) the return probability $R (\bar{R})$
and (f) the transmission probability $T (\bar{T})$.
The model parameters are set to $A=364.5$ nm meV, $B=-686$ nm$^2$ meV, $D=-512$ nm$^2$ meV
$M=-10$ meV.
The Rashba strength at both point contacts is $\alpha_L=\alpha_R=160$nm meV.
The geometric parameters of the system [cf. Appendix \ref{appB}] are set as follows:
the width of QSHI sample in the $y$ direction is $W=300$ nm,
the width of two PCs is $W_{PC}=40$ nm,
the length of two PCs is $L_{PC}=210$ nm
and the length of middle region in between is $L_M=180$ nm.
The results $\bar{R}$ and $\bar{T}$ are averaged over 100 random values of $V_g$.}\label{fig2}
\end{figure}

The conclusion above generally holds
beyond the weak reflection limit.
We provide numerical results of Eq. \eqref{RT}
for a general case in Fig. \ref{fig2}.
The return and transmission probabilities
for fixed and random $V_g$
are shown in Figs. \ref{fig2}(a) and \ref{fig2}(b), respectively.
For a certain $\delta$, the transmission $T$ is
dominated by the AB oscillation with $\Phi_0$
periodicity.
As $\delta$ is averaged out,
both $\bar{T}$ and $\bar{R}$ oscillate with $\Phi_0/2$ periodicity,
which is a clear signature of the AAS scenario.
The transition of the oscillation period from $\Phi_0$ (in $T$) to $\Phi_0/2$ (in $\bar{T}$)
visibly manifests the random gate induced AAS
interference.
Importantly, here the return probability takes the form of $\bar{R}\propto \sin^2\phi$
which vanishes for $\Phi=0$.
It is a direct result of the spin-momentum locking
as elucidated in Sec. \ref{pp}
thus providing a direct evidence of helical edge states.

\section{Lattice model simulation}\label{lm}
Based on the
scattering matrix analysis, one can see
that only the AAS oscillation is stable
for a random $V_g$.
Next, we perform numerical simulation to
show the robustness of the results.
We adopt the Bernevig-Hughes-Zhang model to
describe the QSHI \cite{Bernevig1757}
\begin{equation}
H_{\text{BHZ}}=-Dk^2+Ak_x\tau_x\sigma_z-Ak_y\tau_y+(M-Bk^2)\tau_z,
\end{equation}
where $\sigma_{x,y,z}$ and $\tau_{x,y,z}$ are Pauli matrices acting on
the spin and orbital space, respectively. Here,
$k^2=k_x^2+k_y^2$, and $A, B, D, M$ are the material
parameters. At the PC, the Rashba spin-orbit coupling
is described by $H_R=\alpha(1+\tau_z)(k_x\sigma_y-k_y\sigma_x)$ \cite{rothe2010fingerprint}.
We map the effective model
onto the square lattice through the substitutions
$k^2=2a^{-1}[2-\cos(k_xa)-\cos(k_ya)], k_{x,y}=a^{-1}\sin(k_{x,y}a)$ with $a$
the lattice constant (see Appendix \ref{appB} for details) and calculate the scattering
probabilities using KWANT package \cite{groth2014kwant}.

The geometric parameters of the setup are provided in the caption of
Fig. \ref{fig2} and Appendix \ref{appB}.
The side gate voltage $V_g$ is applied on the upper
edge states, specifically, in the area $3W/8<y<W/2$ of the middle region.
The strengths of Rashba spin-orbit coupling around
two PCs $\alpha_{L,R}$ are set to an equal value.
Consider an electron injected from terminal 1 with
an energy within the bulk gap of the QSHI,
only the edge channels are available for propagation.
The scattering probabilities
for $V_g=0$ is shown in Fig. \ref{fig2}(c),
which resemble those from the edge state analysis in Fig. \ref{fig2}(a). Specifically,
$T$ is dominated by the $\Phi_0$-period oscillation
and $R$ contains multiple periods of oscillation.
Then we consider a random $V_g$ with uniform distribution
$V_g\in[0,\mathcal{W}]$ meV. The averaged scattering probability
$\bar{T}$ and $\bar{R}$ are shown in Fig. \ref{fig2}(d).
Both $\bar{R}$ and $\bar{T}$ exhibit a dominant $\Phi_0/2$-period oscillation,
which manifests a random gate induced AAS effect.
Note that $R$ $(\bar{R})$ vanishes for $\Phi=0$,
indicates the helical spin texture of the edge states.
The transition of the oscillating period is more clearly revealed
by the frequency distribution of
the oscillating pattern in Figs. \ref{fig2}(e)
and \ref{fig2}(f). One can see that there are three
peaks at $1/\phi_0, 2/\phi_0, 3/\phi_0$ in the FFT spectra of $R$ with nearly equal
amplitude for a fixed $V_g$.
The random $V_g$ enhances the $2/\phi_0$ component
in the FFT spectra of $\bar{R}$ while quenches
other frequencies, corresponding to a $\Phi_0/2$-period AAS oscillation. For the transmission
$T$, it is initially dominated by the $1/\phi_0$ frequency corresponding to
the $\Phi_0$-period AB oscillation. After random $V_g$ averaging,
the AB oscillation of $\bar{T}$ quenches,
while the $\Phi_0/2$-period AAS frequency maintains and dominates the transport.

\begin{figure}
\centering
\includegraphics[width=\columnwidth]{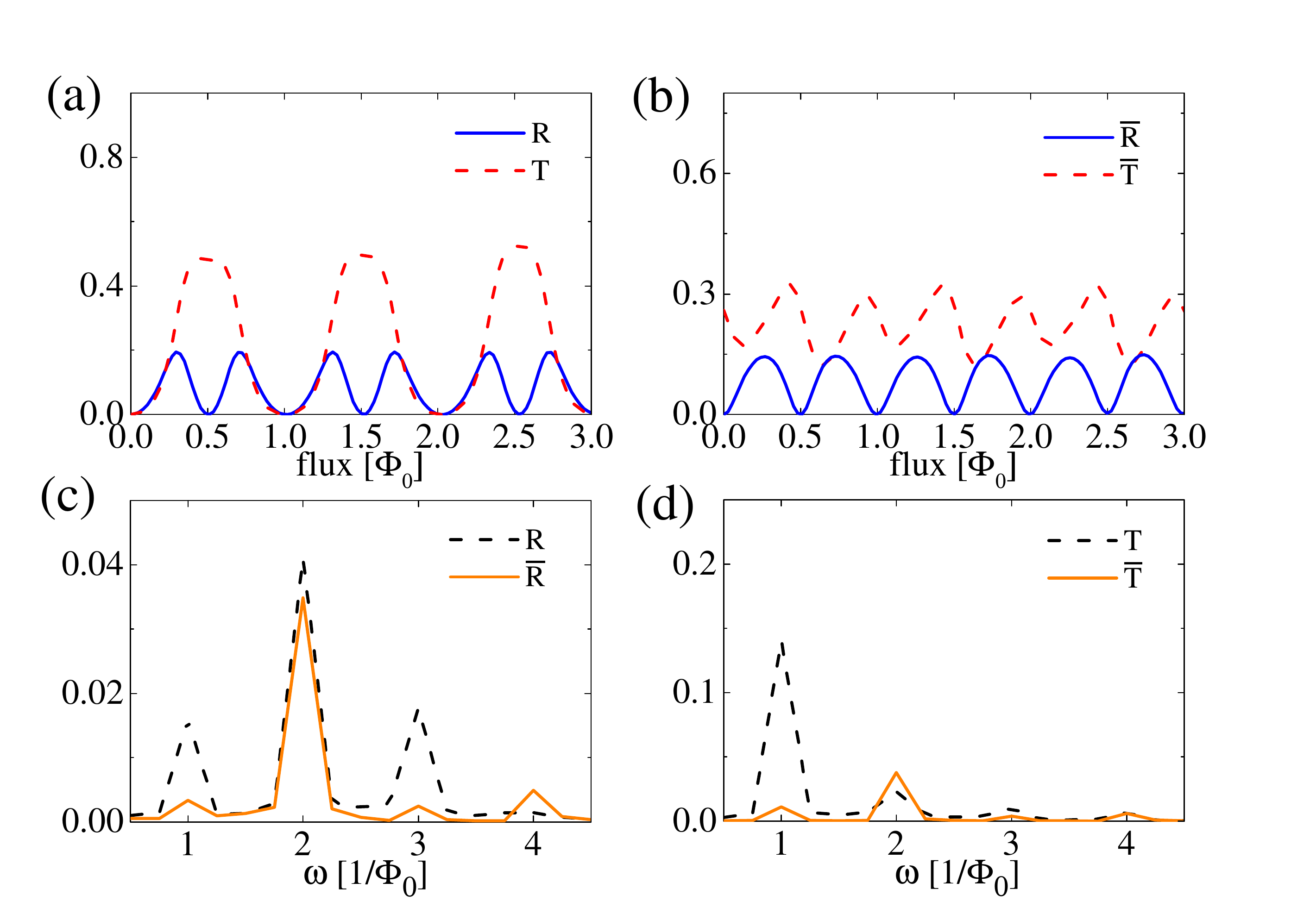}\\
\caption{Oscillation patterns of scattering probabilities
by lattice simulation for (a) $V_g=0$ and (b)
random $V_g$ with $\mathcal{W}=10$meV.
The FFT spectrum of (c) the return probability $R (\bar{R})$
and (d) the transmission probability $T (\bar{T})$.
All parameters are the same as those
in Fig. \ref{fig2} except for $\alpha_R=160$ nm meV.}\label{fig3}
\end{figure}

\begin{figure}
\centering
\includegraphics[width=\columnwidth]{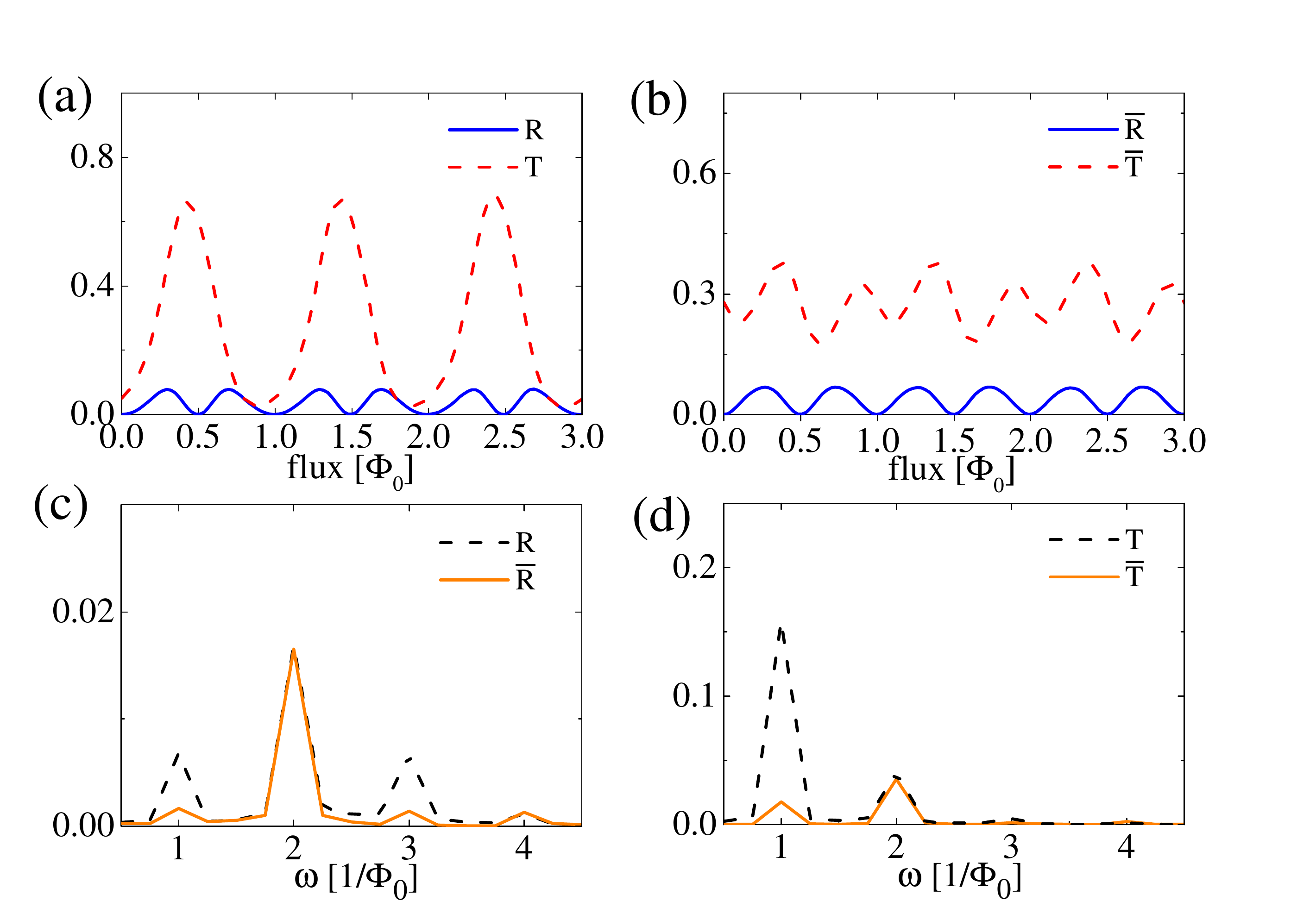}\\
\caption{Oscillation patterns of scattering probabilities
and their FFT spectra.
All parameters are the same as those
in Fig. \ref{fig2} except for
$L_{RPC}=160$ nm.}\label{fig4}
\end{figure}

\begin{figure}
\centering
\includegraphics[width=\columnwidth]{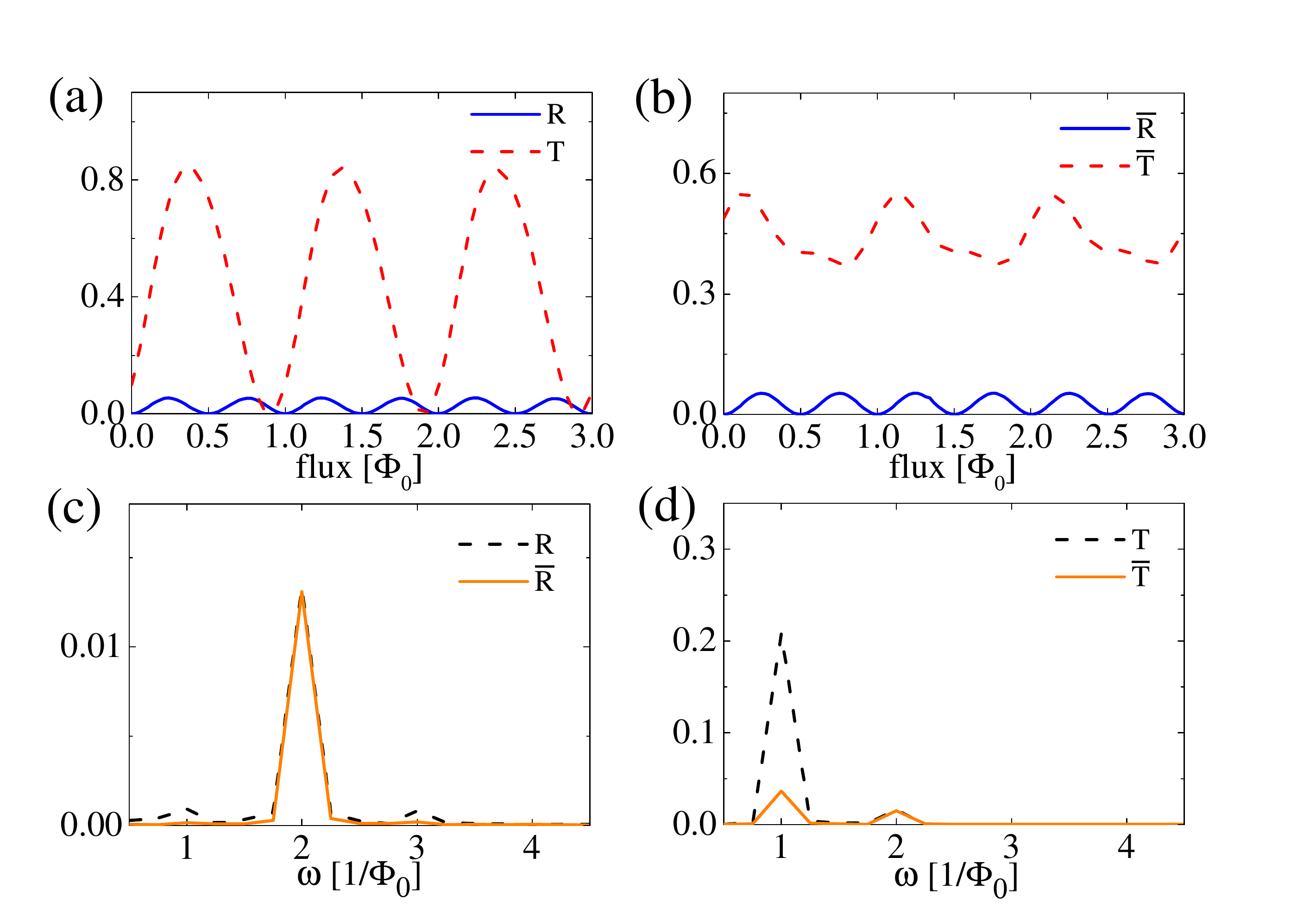}\\
\caption{Oscillation patterns of scattering probabilities
and their FFT spectra.
All parameters are the same as those
in Fig. \ref{fig2} except for a different incident energy of 2 meV.}\label{fig5}
\end{figure}

\begin{figure}
\centering
\includegraphics[width=\columnwidth]{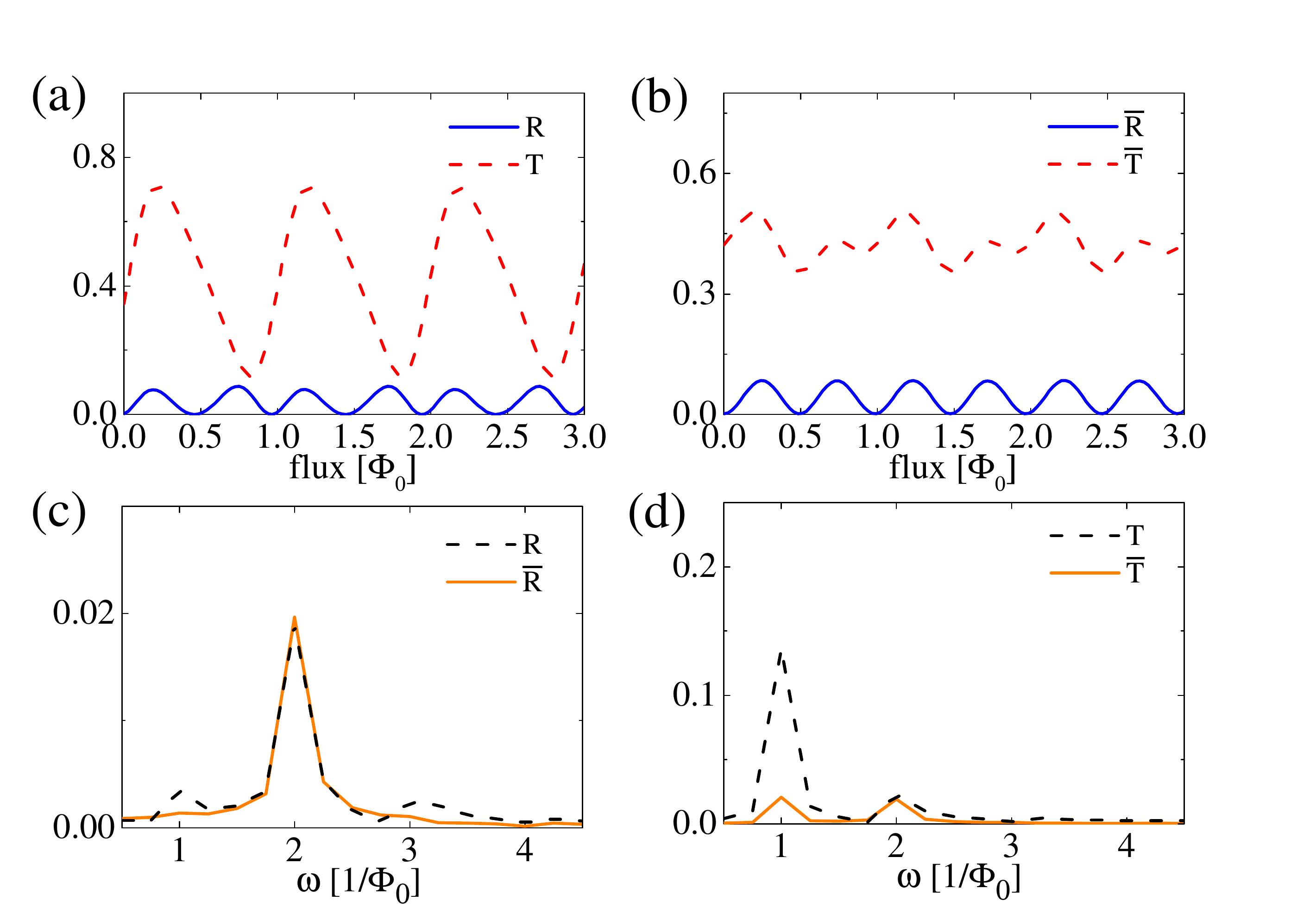}\\
\caption{Oscillation patterns of scattering probabilities
and their FFT spectra with disorder strength 5 meV.
All parameters are the same as those
in Fig. \ref{fig2}.}\label{fig6}
\end{figure}
Next we show that the random gate induced AAS effect generally holds for different systemic
parameters and is robust against nonmagnetic disorder.
First, we change the Rashba coefficient $\alpha_R$ at the right PC.
The scattering probabilities $T, R$ ($\bar{T}, \bar{R}$) without (with) $V_g$ averaging
are shown in Fig. \ref{fig3}. One can see that
the main results still hold, i.e.,
the random $V_g$ quenches the $\Phi_0$-period oscillation and
leads to a dominant $\Phi_0/2$-period oscillation as shown in Figs. \ref{fig3}(a) and \ref{fig3}(b).
There are three peaks in the frequency domain of $R$,
and only a single peak at $2/\phi_0$ survives in $\bar{R}$ after $V_g$ averaging; see Fig. \ref{fig3}(c).
Accordingly, the $\Phi_0/2$-period AAS oscillation overwhelms the $\Phi_0$-period AB oscillation
in the transmission probability $\bar{T}$ as shown in Fig. \ref{fig3} (d).
Similar results also hold as one varies the length $L_{RPC}$ of the right PC
as shown in Fig. \ref{fig4}.
Fig. \ref{fig5} shows the similar results with a different incident energy.
The $\Phi_0/2$-period oscillation dominates the return probability ($R$ and $\bar{R}$)
before and after $V_g$ averaging.
The $\Phi_0$-period oscillation of the transmission probability
is strongly suppressed by the random $V_g$, while the $\Phi_0/2$ component
remains unaffected. We also show in Fig. \ref{fig6} that
our results are robust against disorder. Since the main results
are stabilized by the time-reversal symmetry,
the modification of the sample details will not change the qualitative results.
From Figs. \ref{fig3}-\ref{fig6}, one can see that
the averaged return probability $\bar{R}(\Phi=0)=0$ generally holds for various
sample parameters, indicating the universality
of the predicted signal of the helical edge states.
\section{Trivial edge states}\label{te}
\begin{figure}[htbp]
\centering
{
\begin{minipage}[t]{0.44\columnwidth}
\centering
\includegraphics[width=\columnwidth]{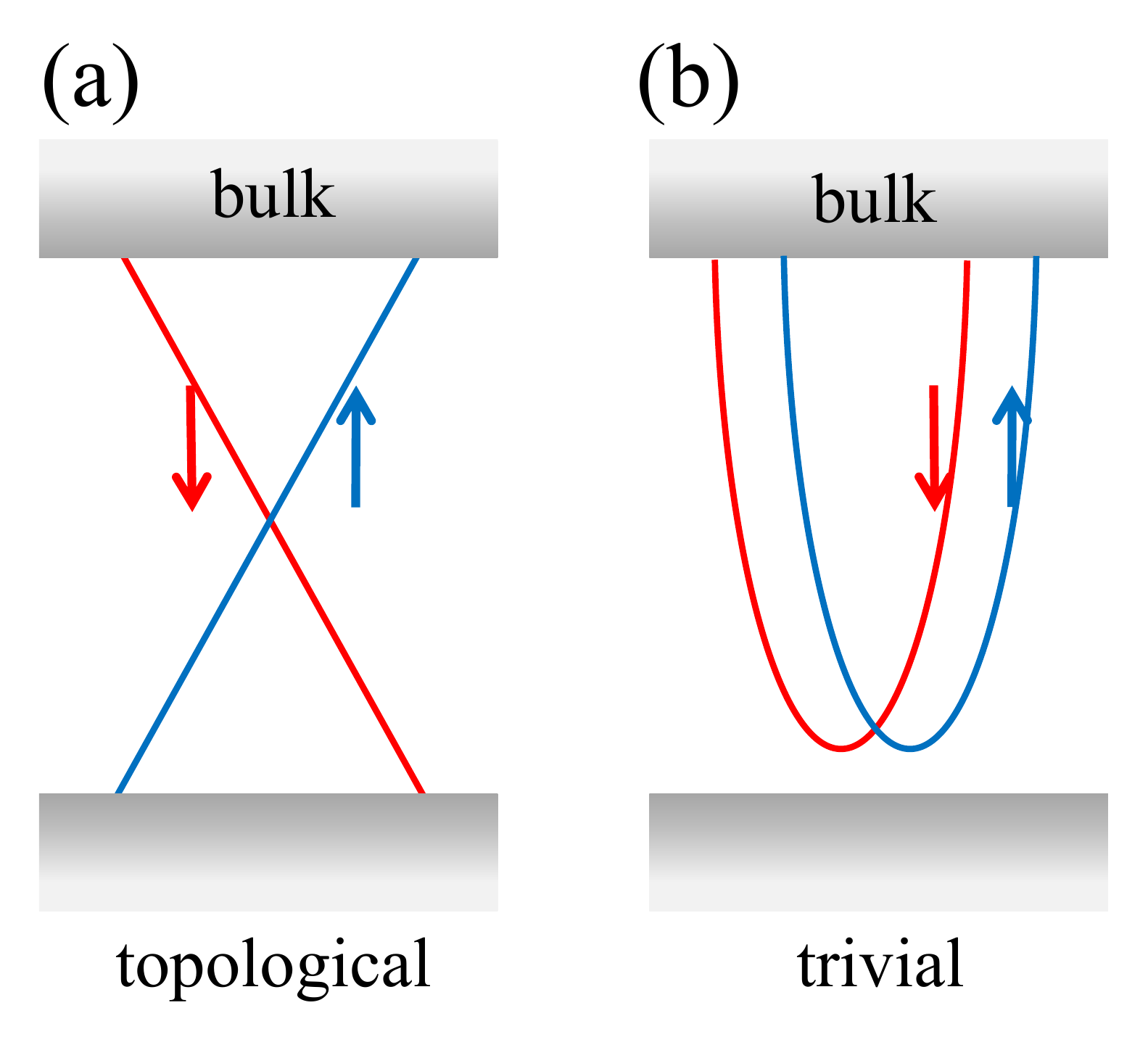}
\end{minipage}%
}
{
\begin{minipage}[t]{0.5\columnwidth}
\centering
\includegraphics[width=\columnwidth]{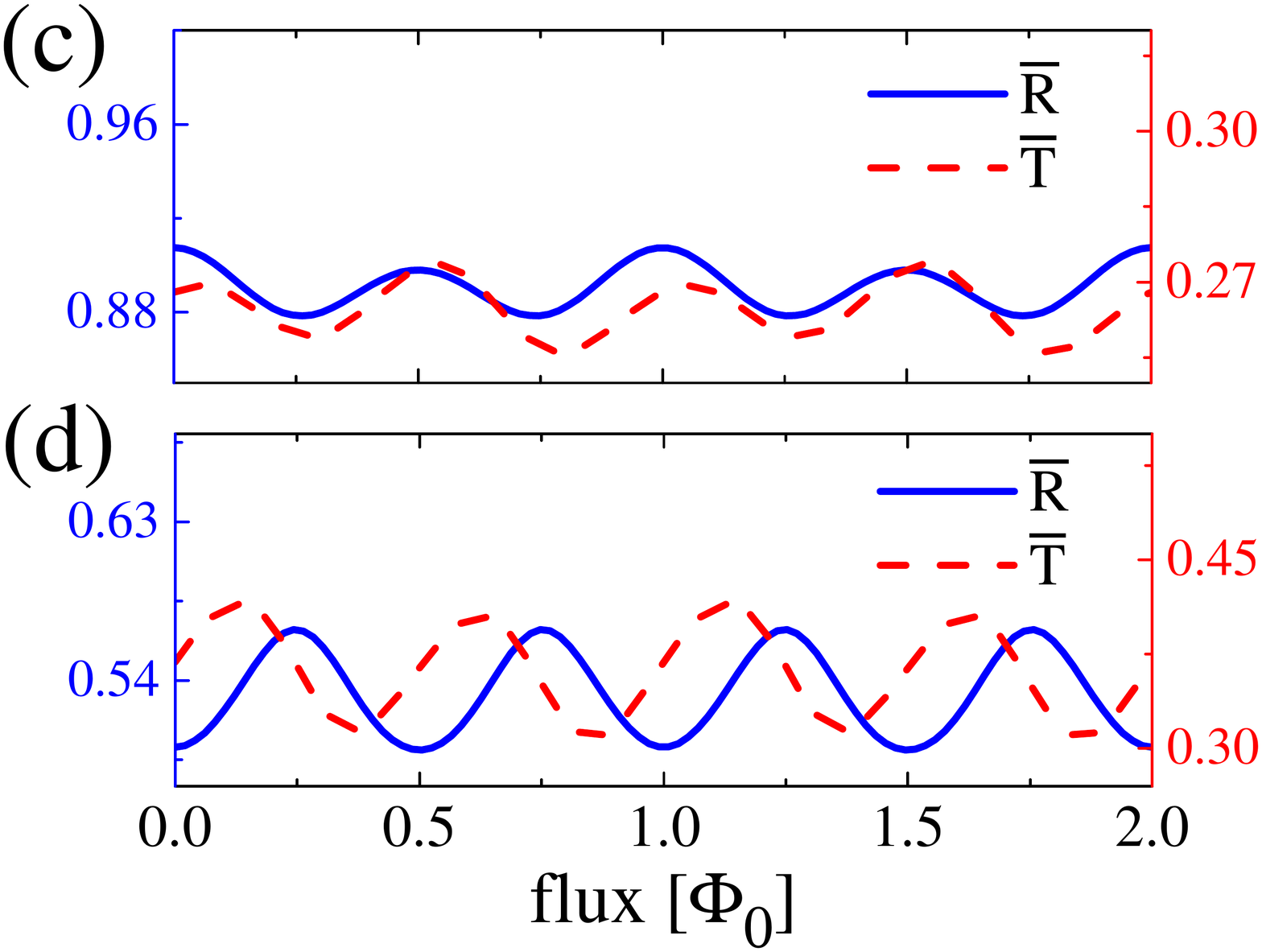}
\end{minipage}%
}
\centering
\caption{Sketch of the band structures of (a) topological edge states and
(b) trivial edge states. The scattering probabilities of the trivial
edge states with (c) $\tilde{\alpha}=0.5$ nm meV and (d) $\tilde{\alpha}=4$ nm meV.}
\label{fig7}
\end{figure}
Very recently, trivial edge states have been reported in various QSHI candidates,
which contribute similar conduction signature to that by the helical
edge states \cite{Nguyen16prl,Nichele16njp,Mueller17prb,Sazgari19prb,Vries18prl,Vries19prr}.
Thus, it is difficult to discriminate the two kinds of edge
states with totally different physical origins.
The main difference between the helical and trivial edge states
can be seen in Fig. \ref{fig7}(a) and \ref{fig7}(b).
The helical edge states originate from
the band topology of QSHI,
which cannot be realized in any one-dimensional nanowire.
In contrast, the trivial edge states originate from
the surface band bending without topological reasons \cite{Nichele16njp},
which resembles those of a nanowire with Rashba spin-orbit coupling \cite{Haidekker20prl},
as shown in Fig. \ref{fig7}(b). The trivial edge
states differ from the helical ones by a doubling
of the transport channel, in which backscattering
with spin conservation becomes possible.
Therefore, the existence of backscattering
or not provides a distinctive signal for the identification
of helical and trivial edge states, which
can be probed in the oscillation pattern
of the return probability.

We simulate the trivial edge states by a one-dimensional
nanowire with Rashba spin-orbit coupling
described by the Hamiltonian $H_{\text{Tri}}=k_x^2/(2m)
+\tilde{\alpha}k_x\sigma_y$.
The AB and AAS effects are studied numerically
in the interferometer
composed of two nanowires coupling at
two points based on the lattice model.
A magnetic flux threads the closed loop and a gate voltage is imposed to the upper
edge like that in the QSHI model. At the coupling bond between
two nanowires, the Rashba spin-orbit coupling $\tilde{\alpha}$ during
lateral hopping is included. The oscillation patterns
for $\bar{T}$ and $\bar{R}$ after $V_g$ averaging
are plotted in Figs. \ref{fig7}(c) and \ref{fig7}(d).
One can see that the averaged scattering probabilities
exhibit a $\Phi_0/2$-period oscillation, indicating
an AAS effect dominated scenario. Specifically,
for a small Rashba coefficient, the return probability
can be captured by $\bar{R}=-\mathcal{R}\sin^2(2\pi\Phi/\Phi_0)+\mathcal{R}_0$; see Fig. \ref{fig7}(c).
A small magnetic field results in a decrease of $\bar{R}$,
showing the weak localization effect.
For a large Rashba coefficient, the oscillation
acquires an extra $\pi$ phase shift, i.e.,
$\bar{R}=\mathcal{R}\sin^2(2\pi\Phi/\Phi_0)+\mathcal{R}_0$ as shown in Fig. \ref{fig7}(d),
which is the weak antilocalization effect.
The feature of $\bar{R}$ pattern for the trivial edge states
provides an effective way to discriminate them from
helical edge states. As the oscillation of $\bar{R}$
shows a weak localization behavior as that in Fig. \ref{fig7}(c),
trivial edge states are confirmed.
Importantly,
in both weak localization and antilocalization regimes
of the trivial edge states, the return probability is always finite
for $\Phi=0$, in start contrast to
the helical edge states, thus providing
a discriminative signal of the helical and trivial edge states.

\section{Discussion and Summary}\label{sd}

Some remarks are made below about the experimental implementation of our proposal.
The main ingredient of the interferometer is the PC structure
introducing inter-edge coupling
which can be achieved by state-of-the-art fabrication techniques \cite{strunz2020interacting}.
Then two PCs make up an interferometer giving rise to coherent oscillation
of the scattering probabilities
with varying magnetic field.
The distinctive signature of helical and trivial edge states
can be extracted from the oscillation pattern of the
return probability $\bar{R}$, which is
directly related to the differential conductance by
$G=\frac{e^2}{h}(1-\bar{R})$. Specifically, as the oscillation
of $\bar{R}$ exhibits a weak localization scenario,
trivial edge states can be identified. If
$\bar{R}$ shows a weak antilocalization oscillation,
the helical and trivial edge states
can be discriminated by zero and finite $\bar{R}$ at $\Phi=0$, respectively.
Note that the precondition for such a judgement is
the occurrence of coherent oscillation. In our proposal,
the absence of return probability should
originate from a spin-texture resolved
interference effect rather than other trivial physical reasons.
The $V_g$ averaged return probability $\bar{R}|_{\Phi=0}=0$ by AAS interference further
reveals the robustness and universality of such a signal,
which excludes any accidental result $R|_{\Phi=0}=0$
by fine tuning the parameters.

To summarize, we investigate quantum coherent transport
through an AB interferometer constructed by the edge states
of a QSHI. By applying a random gate to the edge channel,
the $\Phi_0$-period AB oscillation quenches and the $\Phi_0/2$-period
AAS oscillation dominates the transport.
Such random gate induced AAS oscillation differs
from the conventional scenario of disorder averaging.
The helical spin texture of topological edge states results in
an AAS oscillation of the weak antilocalization type
and more importantly the absence of return probability at zero magnetic field.
In contrast, the AAS oscillation for the trivial edge states
can be of either weak localization or antilocalization type,
both having nonzero return probability.
Therefore, our proposal provides an effective way
for the discrimination between helical and trivial
edge states through spin resolved interference effect.
Such a signal is protected by time-reversal symmetry,
so that it generally holds without fine tuning of the system
and is robust against disorder.

\begin{acknowledgments}
This work was supported by the National
Natural Science Foundation of
China under Grant No. 12074172 (W.C.), No. 11674160 and
No. 11974168 (L.S.), the startup
grant at Nanjing University (W.C.), the State
Key Program for Basic Researches of China
under Grants No. 2017YFA0303203 (D.Y.X.)
and the Excellent Programme at Nanjing University.
\end{acknowledgments}

\begin{appendix}

\section{Lattice model for numerical calculation}\label{appB}
In this Appendix, we elucidate the parameters for the interferometer sketched in Fig. 1(a).
The hard-wall boundary in the $y$ direction is described by the following function
\begin{widetext}
\begin{equation}
|y(x)|=\begin{cases}
\left( W-(W-W_{PC}e^{-(x+L_M/2+L_{PC})^2/30^2}) \right)/2, & -L<x<-(L_M/2+L_{PC}),\\
W_{PC}/2, & -(L_M/2+L_{PC}) \leq x \leq -L_M/2,\\
\left( W-(W-W_{PC}e^{-(x+L_M/2)^2/30^2}) \right)/2, & -L_M/2<x<0,\\
\left( W-(W-W_{PC}e^{-(x-L_M/2)^2/30^2}) \right)/2, & 0 \leq x<L_M/2,\\
W_{PC}/2, & L_M/2 \leq x \leq (L_M/2+L_{PC}),\\
\left( W-(W-W_{PC}e^{-(x-L_M/2-L_{PC})^2/30^2}) \right)/2, & (L_M/2+L_{PC})<x<L,
\end{cases}
\end{equation}
\end{widetext}
where $W$ is the width of the lead, $W_{PC}$ and $L_{PC}$ are the width and length of PC.

The whole devices can be described by Hamiltonian $H = H_{QSH}+H_R$, where $H_{QSH}$ is the Hamiltonian of QSHI and $H_R$ is the Hamiltonian of SOC. Here, we use HgTe/CdTe quantum wells in our proposal, which can be described by Bernevig-Hughes-Zhang model as
\begin{equation*}
\begin{aligned}
  H_{QSH} &= -Dk^2 + Ak_x\tau_x\sigma_z - Ak_y\tau_y + (M - Bk^2)\tau_z\\
  &= \left(
      \begin{array}{cccc}
            D_k+M_k & Ak_+ &0 &0\\
            Ak_- & D_k-M_k &0 &0 \\
            0 & 0 &D_k+M_k &-Ak_-\\
            0 & 0 &-Ak_+ &D_k-M_k \\
      \end{array}
    \right),
\end{aligned}
\end{equation*}
where $\sigma_{x,y,z}$ and $\tau_{x,y,z}$ are Pauli matrices of spin and orbital respectively. $k_\pm=k_x \pm ik_y$, $k^2=k_x^2+k_y^2$, $D_k=-Dk^2$, and $M_k=M-Bk^2$. A, B, D and M are the material parameters which can be controlled by experiment. The Hamiltonian of SOC is
\begin{equation*}
\begin{aligned}
  H_R &= 0.5\alpha(1 + \tau_z)(k_x\sigma_y - k_y\sigma_x)\\
  &= \left(
      \begin{array}{cccc}
            0 & 0 &-i\alpha k_- &0\\
            0 & 0 &0 &0\\
            i\alpha k_+ & 0 &0 &0\\
            0 & 0 &0 &0\\
      \end{array}
    \right),
\end{aligned}
\end{equation*}
where $\alpha$ is Rashba coefficient.

In order to run numerical calculations, we use a square lattice model for whole system by discretizing the continuous effective hamiltonian $H$. By using $k^2=2a^{-2}[2-\cos(k_xa)-\cos(k_ya)]$, $k_x=a^{-1}\sin(k_xa)$, $k_y=a^{-1}\sin(k_ya)$, we can derive the Hamiltonian to real space. The lattice Hamiltonian is
\begin{equation}\label{Hamiltonian}
\begin{split}
H = \sum_{i=1}c_i^{\dagger}H_{ii}c_i + \sum_{i=1}c_i^{\dagger}H_{i,i+a_x}c_{i+a_x} \\
+ \sum_{i=1}c_i^{\dagger}H_{i,i+a_y}c_{i+a_y} + H.C.,
\end{split}
\end{equation}
where $c_i=(c_{s,\uparrow,i},c_{p,\uparrow,i},c_{s,\downarrow,i},c_{p,\downarrow,i})$ are the annihilate operators of electron with spin up and spin down in s and p orbits at site i. The $H_{ii}$, $H_{i,i+a_x}$ and $H_{i,i+a_y}$ are 4 $\times$ 4 Hamiltonians,
\begin{equation}
\begin{split}
\begin{aligned}
  H_{ii}&=-\frac{4D}{a^2} - \frac{4B}{a^2}\tau_z + M\tau_z, \\
  H_{i,i+a_x}&=\frac{D+B\tau_z}{a^2} + \frac{A\tau_x\sigma_z}{2ia}+\frac{\alpha(1+\tau_z)\sigma_y}{4ia}, \\
  H_{i,i+a_y}&=\frac{D+B\tau_z}{a^2} + \frac{iA\tau_y}{2a} + \frac{i\alpha(1+\tau_z)\sigma_x}{4a},
\end{aligned}
\end{split}
\end{equation}
where lattice constant is $a=3$nm.

\end{appendix}


%

\end{document}